\documentclass[twocolumn,showpacs,preprintnumbers,amsmath,amssymb,aps]{revtex4-1}
\pdfoutput=1

\usepackage{graphicx}% Include figure files
\usepackage{dcolumn}% Align table columns on decimal point
\usepackage{bm}% bold math
\usepackage{color}

\usepackage[normalem]{ulem}
\usepackage{amssymb,amsmath}
\usepackage{enumitem}

\newcommand{\ti}[1]{\tilde{#1}}
\newcommand{\mm}{\bar{\mu}_s}

\begin{document}
\title{Avalanche precursors in a frictional model}

\author{Axelle Amon} \email{axelle.amon@univ-rennes1.fr}
\affiliation{Institut de Physique de Rennes, UMR UR1-CNRS 6251,
  Universit\'e de Rennes 1, Campus de Beaulieu, F-35042 RENNES cedex,
  France}
\author{Baptiste Blanc}\altaffiliation{present address:}

\author{Jean-Christophe G\'eminard}
\affiliation{Universit\'e de Lyon, Laboratoire de Physique, Ecole
  Normale Sup\'erieure, CNRS, 46 All\'ee d'Italie, F-69364 Lyon cedex 07, France}
\date{Received: date / Revised version: date}

\begin{abstract}
We present a one-dimensional numerical model based on elastically
coupled sliders on a frictional incline of variable tilt. This very
simple approach makes possible to study the precursors to the
avalanche and to provide a rationalization of different features that
have been observed in experiments. We provide a statistical
description of the model leading to master equations describing the
state of the system as a function of the angle of inclination. Our
central results are the reproduction of large-scale regular events
preceding the avalanche, on the one hand, and an analytical approach
providing an internal threshold for the outbreak of rearrangements
before the avalanche in the system, on the other hand.
\end{abstract}

\pacs{45.70.Ht,45.70.-n,46.55.+d,45.05.+x}

\maketitle

%%%%%%%%%%%%%%%%%%%%%%%%%%%%%%%%%%%%%%%%%%%%%%%%%%%%%%%%%%%%%%%%%%%%%%%%%%%%%%%%%

\section{Introduction}

Identifying precursors to avalanches has inspired numerous works with
the hope of being able to detect catastrophic events before they
occur. A physicist approach consists in simplifying the system in
order to single out the fundamental mechanisms underlying the
phenomenon. For example, models based on cellular automaton have been
proposed to study avalanches~\cite{Bak1987}. From an experimental
point of view, a model set-up to study the behavior of a granular
material before an avalanche takes place consists in the progressive
inclination of a box filled with a granular
material~\cite{Nerone2003,Kiesgen2012,Amon2013,Duranteau2013,Gravish2014}. In
such experiments, it has been evidenced that the response of the
system consists in the superposition of two different behaviors: on
the one hand, small rearrangements implying only a few number of grains
occuring without any obvious synchronisation across the system and, on
the other hand, large correlated events implying a large fraction of
the system. Those last events have been shown to emerge from an angle
at about half the avalanche angle~\cite{Amon2013} and then to occur at
regular angle increments as the inclination increases until the
destabilization of the pile~\cite{Nerone2003,Kiesgen2012,Amon2013}. To
our knowledge, no explanation of the regularity of those pre-avalanche
events exists in the litterature, nor any numerical observation of the
phenomenon~\cite{Staron2002,Staron2006,Welker2011}. Yet, such
micro-envents have been observed in different loading
configurations~\cite{Nerone2003,Nguyen2011,LeBouil2014a} and are
reminiscent of precursors observed in studies of the onset of
frictional sliding~\cite{Rubinstein2007}.

Here, we present a one-dimensional frictional model that reproduces
most of the features reported in the experiments. In particular, the
regular micro-ruptures are reproduced and an interpretation of the
origin of the regularity of the phenomenon as a stick-slip response is
given. Synchronization of this periodic response in large, disordered,
systems is exhibited by incorporating a global coupling in the
model. The physical origin of this coupling is discussed. The
simplicity of our model makes possible to provide master equations to describe
the evolution state of the system as the inclination is
increased. This analytical approach reveals an internal
angle signing an intensification of the process.

The article is structured as follow: part II is devoted to the
description of the numerical model. Part III presents the results
obtained for three typical sets of parameters: a small system without
global coupling, a large system without global coupling and finally a
large system with global coupling. Part IV details a statistical
approach to rationalize our observation. The last part discusses the
understanding of the pre-avalanche behavior that can be obtained from
our approach and compare our results to previous studies from the
litterature.

%%%%%%%%%%%%%%%%%%%%%%%%%%%%%%%%%%%%%%%%%%%%%%%%%%%%%%%%%%%%%%%%%%%%%%%%%%%%%%%%%

\section{Numerical model}\label{sec:method}
\subsection{Description}\label{sec:method_1}

The system under study is very close in spirit to the one-dimensional
uniform Burridge-Knopoff model~\cite{Burridge1967}. This seminal work
has inspired numerous studies aiming at understanding the dynamics of
earthquake faults using spring-blocks
models~\cite{Burridge1967,Carlson1989,Carlson1994}. Similar models are
also studied to understand the onset of frictional sliding between two
interfaces in tribological studies~\cite{Braun2009,Maegawa2010}. In
our case we build on a model which has been previously proposed to
study the effects of minute temperature changes on the stability of
granular materials~\cite{Blanc2011,Blanc2014}. The main features of
the present study compared to previous ones are the following: (i) we
consider an homogeneous gravitational loading due to the progressive
inclination of the blocks; (ii) we are interested in the evolution of
the system from an initial preparation to the critical state which
immediately preceed the avalanche; (iii) we propose an implementation
of some long-range coupling in the system. We discuss the physical
origin of this global coupling in section~\ref{sec:discussion}.

\begin{figure}[ht!]
  \centering
  \includegraphics[width=.9\columnwidth]{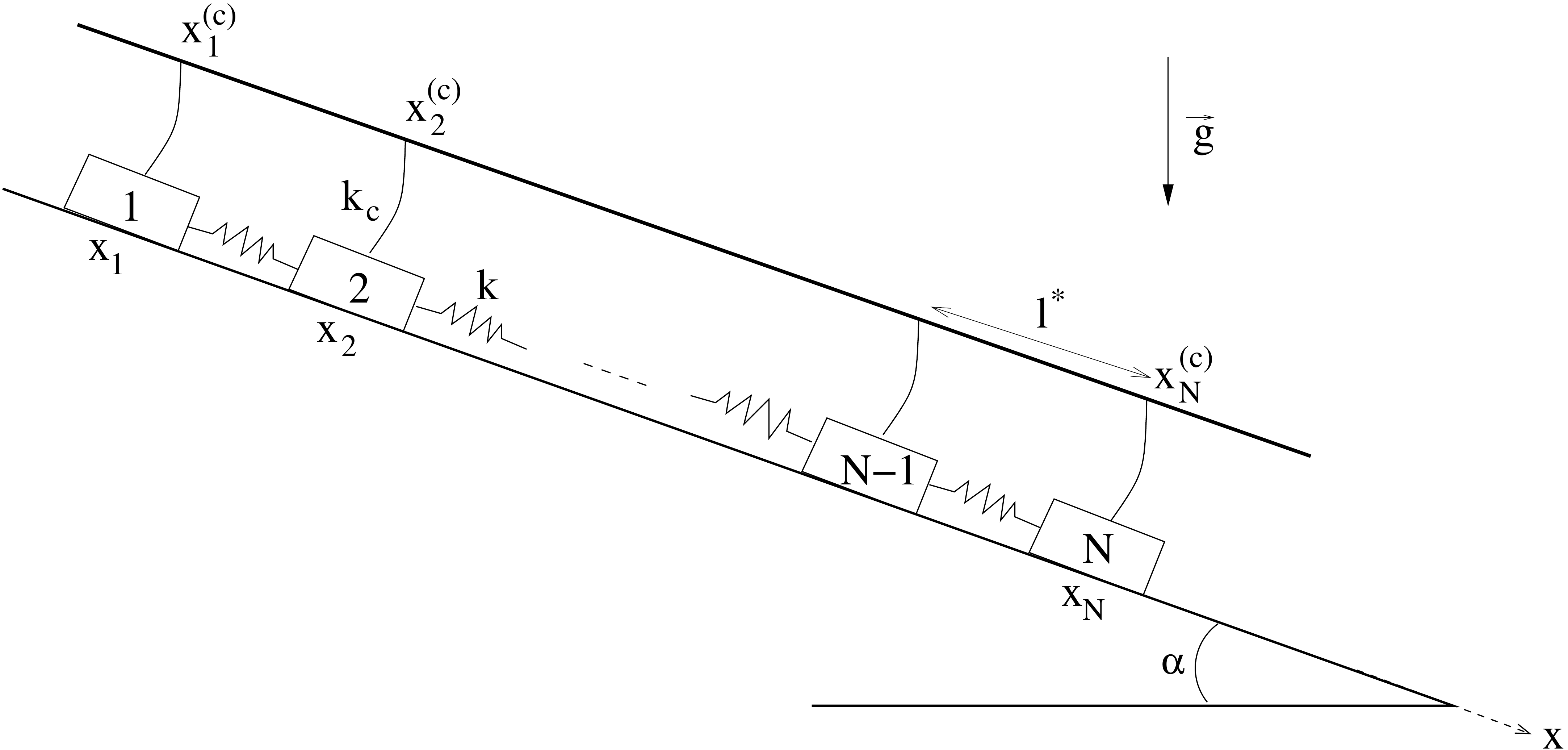}
  \caption{Sketch of the system under study.}
  \label{fig:schema}
\end{figure}

We consider $N$ identical frictional sliders of mass $m$ lying on a
rigid incline making the angle $\alpha$ with the horizontal
(Fig.~\ref{fig:schema}).  The sliders are connected to one another by
linear springs of stiffness $k$.  In addition, each slider is
connected to a rigid upper plate by a cantilever spring of stiffness
$k_c$. The upper plate insures a global coupling between the sliders.
The springs as well as the upper plate are massless.  In this
configuration, the sliders are subjected to the elastic forces due to
the springs, to their own weight $m g$, with $g$ the acceleration due
to gravity, and to the reaction force from the incline which includes
a frictional force. The notations used in the following are shown in
Fig.~\ref{fig:schema}.

One of the fundamental differences between the system under study and
the usual Burridge-Knopoff (BK) models is found in the loading
process. Indeed, we focus on the response of the system to a
progressive increase of the inclination angle $\alpha$, whereas the
system is usually loaded by pulling it at one extremity. We prepare
the system at horizontal, and then progressively increase the tilt in
a quasi-static way until all the sliders descend the slope, which
corresponds to the final avalanche.  We are interested in the temporal
and spatial distribution of the rearrangements that lead to the
avalanche. In particular, we want to study how the distribution of the
static frictional forces exerted on the sliders are modified from an
initially uniform preparation because of the progressive loading.

Another difference is found in the modeling of the frictional contact.
Whereas, in most of the BK-models velocity weakening of the friction
force is introduced, we here characterize the frictional contact
between the incline and the sliders by static and dynamic frictional
coefficients.  However, we consider that, due to the heterogeneity of
the local properties of the incline surface, the static coefficient
$\mu_{s,n}$ takes a different value for each of the individual
sliders, indexed by $n$ (from 1 to $N$, Fig.~\ref{fig:schema}).  By
constrast, because the sliders in motion see average properties of the
incline surface, we consider a single value $\mu_d$ of the dynamical
frictional coefficient for all the sliders.  Note that, in addition,
we consider that $\mu_d < \mu_{s,n}~(\forall n$).  Thus, in summary,
we write that the slider $n$ starts moving if
\begin{eqnarray}
\left| f_{n+1 \rightarrow n} + f_{n-1 \rightarrow n} + f_{C \rightarrow n} + m g\,\sin \alpha \right| \nonumber\\
  > \mu_{s,n}\,m g\,\cos{\alpha}
  \label{eq:stability}
\end{eqnarray}
where $f_{n+1 \rightarrow n}$ and $f_{n-1 \rightarrow n}$ are the elastic forces due to the neighbor sliders $n-1$ and $n+1$, and $f_{C \rightarrow n}$ the elastic force due to the cantilever spring that connects it to the upper plate.
When the slider $n$ is in motion with the velocity $\dot{x}_n$, the frictional force exerted by the incline is
\begin{equation}
f_{d,n} = - \mu_d\,m g\,S(\dot{x}_n)\,\cos{\alpha}
\label{eq:dynamics}
\end{equation}
where $S$ denotes the sign function [$S(u)=1$ if $u>0$ and $S(u)=-1$ if $u<0$].

\subsection{Dimensionless set of equations}
\label{sec:method_2}

The dynamics of the system is characterized by the timescale $\tau_{dyn} = \sqrt{m/k}$.
We can then introduce the dimensionless time $\ti{t} = t/\tau_{dyn}$ and position $\ti{x} = x/g\tau^2_{dyn}$.
We then denote $\tilde{l}$ the dimensionless natural length of the springs that link the sliders to one another and define $\xi \equiv k_c/k$, the ratio of the stiffnesses of the two kinds of springs introduced previously.

Using these latter dimensionless variables, we can write the equations governing the dynamics of a slider in motion.
We have (Eq.~\ref{eq:dynamics}):
\begin{align}
\ddot{\ti{x}}_1 = & -[\ti{x}_1 - \ti{x}_2 + \ti{l}] - \xi [\ti{x}_1 -
  \ti{x}_{1}^{(c)}] \nonumber \\
&  - \mu_d S(\dot{\ti{x}}_1) \cos \alpha + \sin \alpha \hspace{5mm}
(n=1) \label{eq:dyn1}\\
\ddot{\ti{x}}_n = & -[2\ti{x}_n - (\ti{x}_{n+1} + \ti{x}_{n-1})] - \xi [\ti{x}_n -
  \ti{x}_{n}^{(c)}] \nonumber \\
&  - \mu_d S(\dot{\ti{x}}_n) \cos \alpha + \sin \alpha \hspace{5mm}
(n \neq 1,N) \label{eq:dyn2}\\
\ddot{\ti{x}}_N = & -[\ti{x}_N - \ti{x}_{N-1} - \ti{l}] - \xi [\ti{x}_N -
  \ti{x}_{N}^{(c)}] \nonumber \\
&  - \mu_d S(\dot{\ti{x}}_N) \cos \alpha + \sin \alpha \hspace{5mm}
(n=N) \label{eq:dyn3}
\end{align}
where
\begin{equation}
\ti{x}_{n}^{(c)} = \frac{1}{N}\sum_{i=1}^N \ti{x}_i + \left(n -
\frac{N+1}{2}\right)\ti{l}^*
\label{eq:bary}
\end{equation}
with $\ti{l}^*$ the dimensionless, constant, distance ($\ti{l}^* \equiv
l^*/g\tau^2_{dyn}$) between the attachment positions $x_{n}^{(c)}$ of the cantilever springs to the upper plate (Fig.~\ref{fig:schema}, refer to next section addressing the preparation of the system). Eq.~(\ref{eq:bary}) reflects the fact that the upper
plate moves with the barycenter of the sliders.

Depending on its position indexed by $n$, a slider, initially at rest, starts moving if the following condition is fulfilled (Eq.~\ref{eq:stability}):
\begin{align}
\left| \ti{x}_2 - \ti{x}_1 - \ti{l} - \xi \left[\ti{x}_1 -
  \ti{x}_{1}^{(c)}\right] + \right. & \left. \sin \alpha \right|  > \mu_{s,1} \cos
\alpha \nonumber \\ &(n=1) \label{eq:slide1}\\
\left| \ti{x}_{n+1} + \ti{x}_{n-1} - 2 \ti{x}_n - \xi \left[\ti{x}_n -
  \ti{x}_{n}^{(c)}\right] + \right. & \left. \sin \alpha \right| > \mu_{s,n} \cos
\alpha  \nonumber \\ &(n \neq 1,N) \label{eq:slide2}\\
\left| \ti{x}_{N-1} - \ti{x}_N + \ti{l} - \xi \left[\ti{x}_N -
  \ti{x}_{N}^{(c)}\right] + \right. & \left. \sin \alpha \right| > \mu_{s,N} \cos
\alpha \nonumber \\ &(n=N) \label{eq:slide3}
\end{align}

\subsection{Numerical method}
\label{sec:method_3}

In this section, we detail the numerical methods, starting with the crucial preparation of the initial state of the system.

\subsubsection{Preparation of the system}

First, the initial set of static frictional coefficients $\mu_{s,n}^{(0)}$ accounting for the contact betweent the slider $n$ and the incline (angle $\alpha = 0$, initially) are drawn at random from a Gaussian distribution:
\begin{equation}
p(\mu_{s}) = \frac{1}{\sqrt{2\pi\sigma_{\mu}^2}} \exp \left[ -
  \frac{\left( \mu_s - \bar{\mu}_s \right)^2}{2 \sigma_{\mu}^2}
  \right] \label{eq_proba_mus}
\end{equation}
where $\bar{\mu}_s$ is the mean value and $\sigma_{\mu}$ the width of
the distribution. However, if a random value $\mu_d$ exceeds $\mu_s$, we draw it again
at random until we have $\mu_d < \mu_s$.

Once the initial set of static frictional coefficients $\mu_{s,n}^{(0)}$ is chosen,
we set the initial positions of the sliders, $\tilde x_n^{(0)}$, insuring the mechanical stability of the initial configuration in absence of global coupling (The cantilever springs are deconnected, i.e. $\xi=0$). To do so, a set of random tangential forces are drawn according to a uniform distribution in $]-\mm,\mm[$. The corresponding positions, $\tilde x_n^{(0)}$, of the sliders are then computed. When considering the behavior of the system in absence of coupling ($\xi=0$), the obtained configuration is our initial condition.
When the behavior of the system with coupling ($\xi\neq0$) is considered, we alter the configuration in the following way.
First, we set $\ti{l}^* \equiv (\tilde x_N^{(0)}-\tilde x_1^{(0)})/(N-1)$, the average distance between the sliders. Then, the coupling is introduced by connecting the cantilever springs to the upper plate (by setting $\xi$ to a non-zero value).
In response, some sliders can loose stability.
If so, the dynamical equations (\ref{eq:dyn1}) to (\ref{eq:dyn3}) are integrated until a
mechanically-stable configuration is reached. The final positions, $\tilde x_n^{(0)}$, of the sliders are then our initial condition.

\subsubsection{Numerical integration}
\label{sec:integration}

From the initial horizontal state ($\alpha = 0$), the tilt angle $\alpha$ is progressively increased.
From the knowledge of the static frictional coefficients $\mu_{s,n}^{(0)}$ and
positions $x_n^{(0)}$, we can determine the value of the angle $\alpha^{(1)}$
leading to a first rearrangement using Eqs.~\eqref{eq:slide1} to \eqref{eq:slide3}.
The dynamical equations  are then solved to obtain the set of new steady positions $x_n^{(1)}$ of the sliders that moved.
To do so, we set the angle $\alpha$ to $\alpha^{(1)}$ and integrate, using a standard velocity Verlet integrator (with a timestep $\Delta t \ll \tau_{dyn}$), the dynamical equations~\eqref{eq:dyn1} to \eqref{eq:dyn3} for all the sliders that enter in motion.
At each integration step, we check if any of the motionless sliders is destabilized by any
displacement of its neighbors.
In case of such event, the integration step is reduced until the initiation of the motion of the corresponding slider.
In the same way, we check if any slider in motion comes to rest and we
adapt the integration step accordingly.
A new value of the static frictional coefficient is drawn at random, according to Eq.~\eqref{eq_proba_mus}, for those sliders that come to rest.
We consider that the rearrangement ends when all the sliders have come back to rest.
The new state of the system is then accounted for by the sets of new steady positions
$x_n^{(1)}$ and static frictional coefficients $\mu_{s,n}^{(1)}$.

From those new initial conditions the procedure can be iterated, leading to a set of destabilization angles $\alpha^{(k)}$ associated to positions $x_n^{(k)}$ and static frictional coefficients $\mu_{s,n}^{(k)}$. The procedure ends when the angle $\alpha$ reaches a critical value, $\alpha_c$, such that all the sliders enter in motion and continuously accelerate downwards. This final angle is, by definition, the avalanche angle.

\section{Numerical results}
\label{sec:results}

In order to get insights in the behavior of the model, we report
numerical results obtained for three typical systems, small and large systems
with or without coupling:
\begin{itemize}[label=,noitemsep]
 \item A. -- $N=10$ sliders without coupling ($\xi = 0$).
 \item B. -- $N=400$ sliders without coupling ($\xi = 0$).
 \item C. -- $N=400$ sliders with coupling ($\xi \neq 0$).
\end{itemize}
In all the simulations we set the parameters to typical values, i.e.
$l=10$, $\mu_d = 0.55$, $\bar{\mu}_s = 0.6$, and $\sigma_{\mu} = 0.01$.

\subsection{Small system, no global coupling}
\label{sec:results_small}

\begin{figure}[ht!]
  \centering
  \includegraphics[width=\columnwidth]{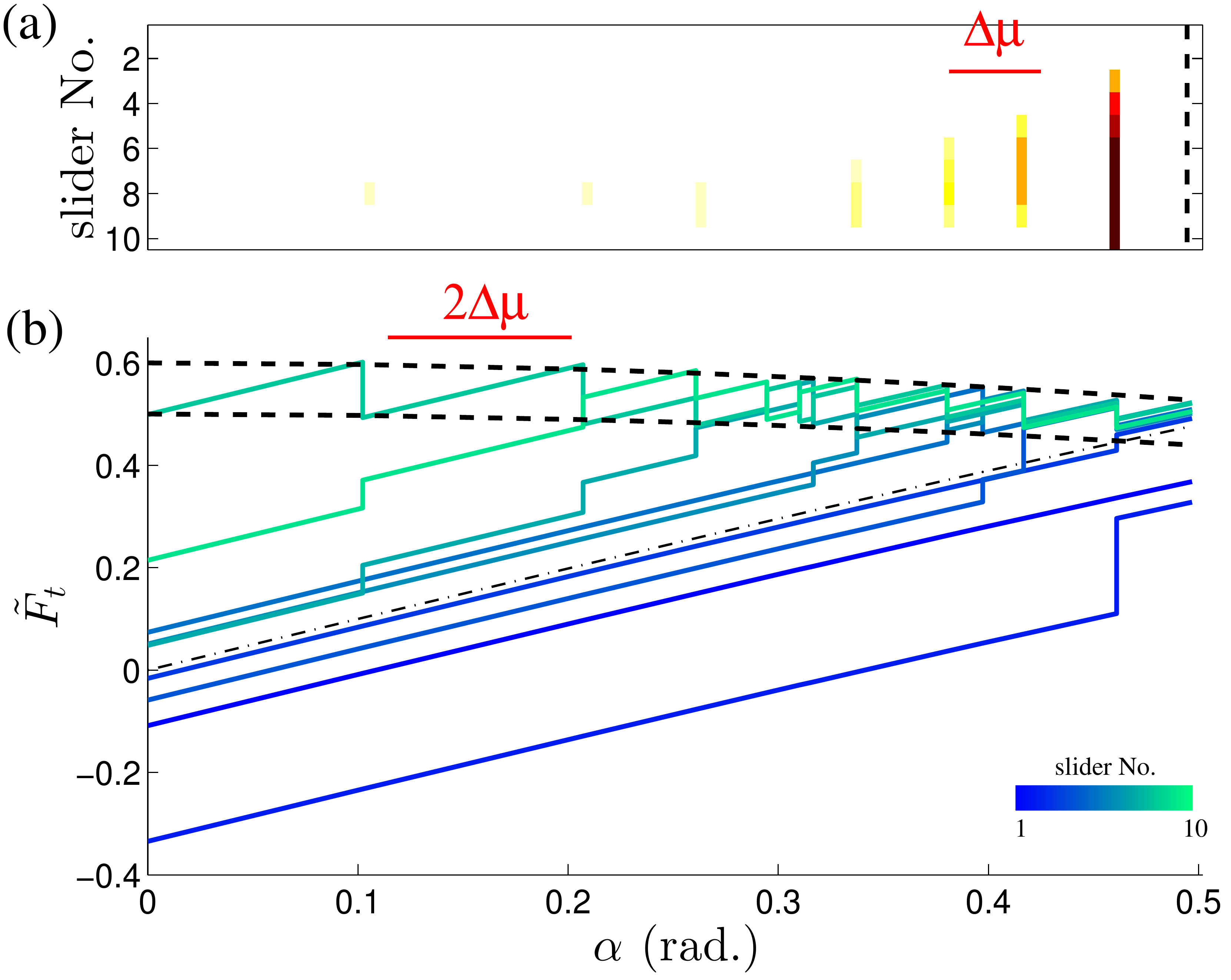}
  \caption{Behavior of a small system without global coupling upon
    inclination -- (a) Hot map of the displacements $\Delta \tilde
    x_i$ vs. angle $\alpha$.  Each horizontal line corresponds to one
    slider.  The darker the points are, the larger is the displacement
    of the slider.  The vertical dashed line corresponds to the
    avalanche angle and $\Delta \mu = \bar{\mu}_s - \mu_d$. (b)
    Tangential force $\tilde F_t$ ($\tilde F_t = \tilde f_{n+1
      \rightarrow n} + \tilde f_{n-1 \rightarrow n} + \tilde f_{C
      \rightarrow n} + \sin \alpha$) vs. angle $\alpha$ (The slider
    number is indicated by the color bar). The upper dashed line is
    the stability limit: $\bar{\mu}_s \cos \alpha$.  The lower dashed
    line is the stability limit $(2\mu_d - \bar{\mu}_s) \cos \alpha$.
    The dash-dotted line: $\sin \alpha$ [$N = 10$, $\xi = 0$, $l=10$,
      $\mu_d = 0.55$, $\bar{\mu}_s = 0.6$, and $\sigma_{\mu} =
      0.01$].}
  \label{fig:small}
\end{figure}

In Fig.~\ref{fig:small} we report results obtained for a small system of $N=10$
sliders in absence of coupling ($\xi = 0$).

We observe in Fig.~\ref{fig:small}a that, far from being uniformly distributed neither
between the sliders nor during the inclination process, the displacements are gathered
in bursts of displacements of adjacent sliders at given angles separated by
quiescent inclination intervals. Moreover, the interval between the inclinations angles $\alpha$
that lead to rearrangments seems to be almost constant.

These features can be understood by considering the evolution of the
tangential component of the force, due to the springs and the weight,
on each of the sliders as function of $\alpha$
(Fig.~\ref{fig:small}b): $\tilde F_t = \tilde f_{n+1 \rightarrow n} +
\tilde f_{n-1 \rightarrow n} + \tilde f_{C \rightarrow n} + \sin
\alpha$. While $\alpha$ is increased, the component of the weight
along the incline increases as $\sin \alpha$ (dash-dotted line).
Starting from the horizontal ($\alpha = 0$), we observe that the first
rearrangment occurs for $\alpha \simeq 0.1$, when the most unstable
slider looses stability.  The displacement of this slider leads to a
drop of the force it is submitted to.  Simultaneously, the tangential
forces exerted on its two direct neighbors present a sudden increase.
At about $\alpha \simeq 0.2$, the same block again looses stability,
and drags downwards one of its neighbors which then also looses
stability and moves.  A further increase of the inclination by a few
degrees leads to another rearrangment that involves three sliders,
thus a larger region.  Sliders, that were unperturbed up-to-now, are
now adjacent to sliders that move, and the tangential forces they are
subjected to exhibit sudden increases, which places them closer to the
instability threshold.  Hence, step by step, the bursts of
reorganizations imply more and more sliders.  Finally, a critical
angle, $\alpha_c \simeq 0.5$, is reached at which all the sliders
destabilize and accelerate down the incline.  Note in
Fig.~\ref{fig:small}b that the tangential forces exerted on the
sliders that rearranged at least once are bounded by two well-defined
values

\subsection{Large system, no global coupling}
\label{sec:results_large}

\begin{figure}[htbp]
  \centering
  \includegraphics[width=\columnwidth]{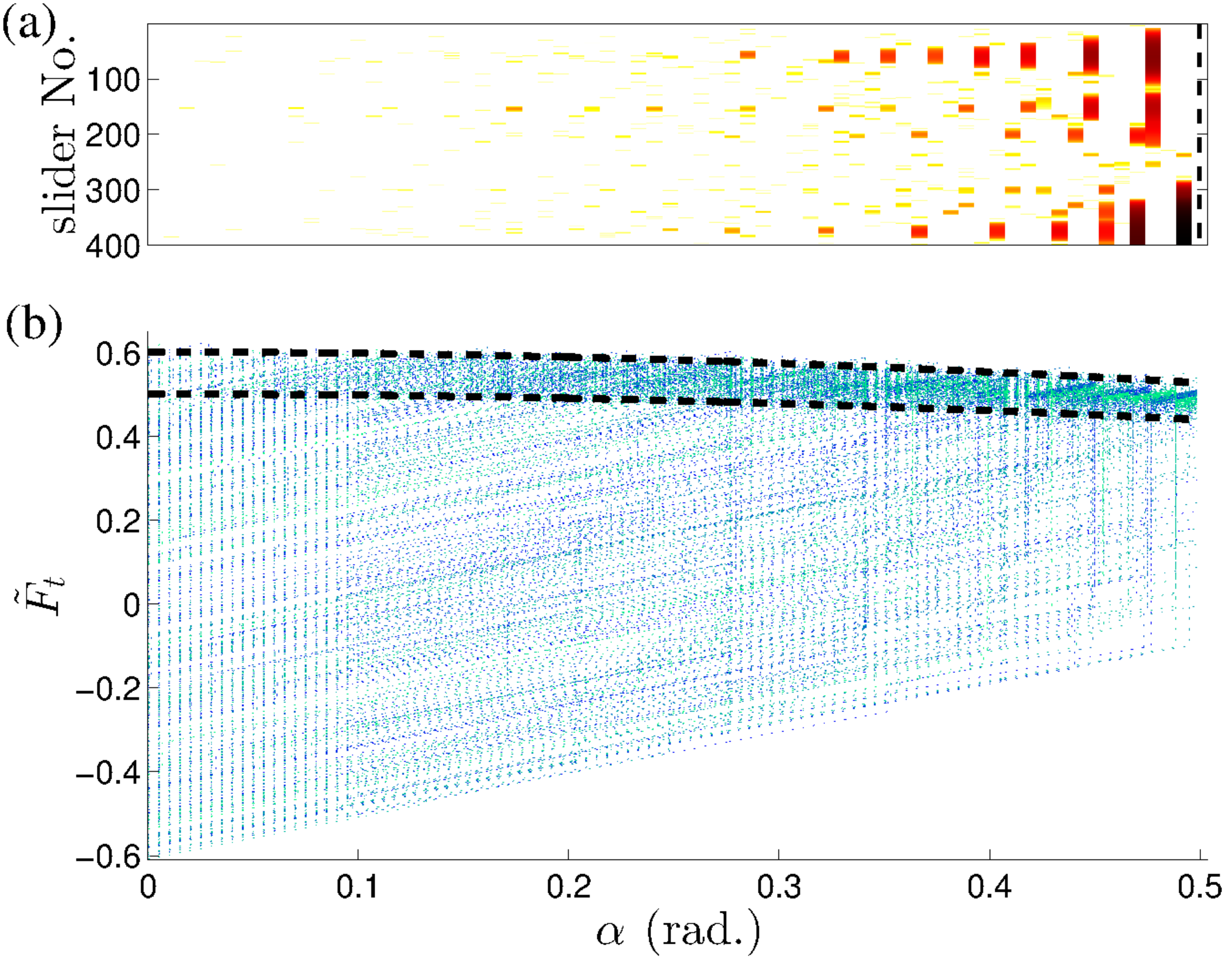}
  \caption{Behavior of a large system without global coupling upon
    inclination -- (a) Hot map of the displacements $\Delta \tilde
    x_i$ vs. angle $\alpha$. (b) Tangential force $\tilde F_t$
    vs. angle $\alpha$ (The slider number is indicated by the color
    bar). The upper dashed line is the stability limit: $\bar{\mu}_s
    \cos \alpha$.  The lower dashed line is the arrest limit $(2\mu_d
    - \bar{\mu}_s) \cos \alpha$ [$N = 400$, $\xi = 0$, $l=10$, $\mu_d
      = 0.55$, $\bar{\mu}_s = 0.6$, and $\sigma_{\mu} = 0.01$].}
  \label{fig:large_no}
\end{figure}

For a large system, we observe the same typical behavior of the system
with the main difference that bursts of rearrangments are observed at
different places in the system (Fig.~\ref{fig:large_no}a).  The
picture is more complex as the rearrangments that occur in different
regions are not synchronized.  In addition, the intervals between the
inclinations angles $\alpha$ are, on average, smaller than previously
observed in a smaller system.  Note also that, upon increasing
inclination, the typical size of the active regions increases, which
leads to their coalescence.

Reporting the tangential components of the forces (Fig.~\ref{fig:large_no}b),
we observe that rearrangements initiate at different locations as the instability
threshold is reached by several sliders early in the inclination process.
Upon further increase of the inclination angle, each of the active regions evolves,
and grows in size, as previously described for the smaller system,
which leads, finally, to the avalanche of the whole system.

\subsection{Large system with global coupling}
\label{sec:results_coupling}

From now on, it is of particular interest to observe the behavior
of the exact same system when a slight global coupling is introduced.
To do so, we set the ratio of the spring constants $k$ and $k_c$ to the small,
but non zero, value $\xi=0.001$.

\begin{figure}[htbp]
  \centering
  \includegraphics[width=\columnwidth]{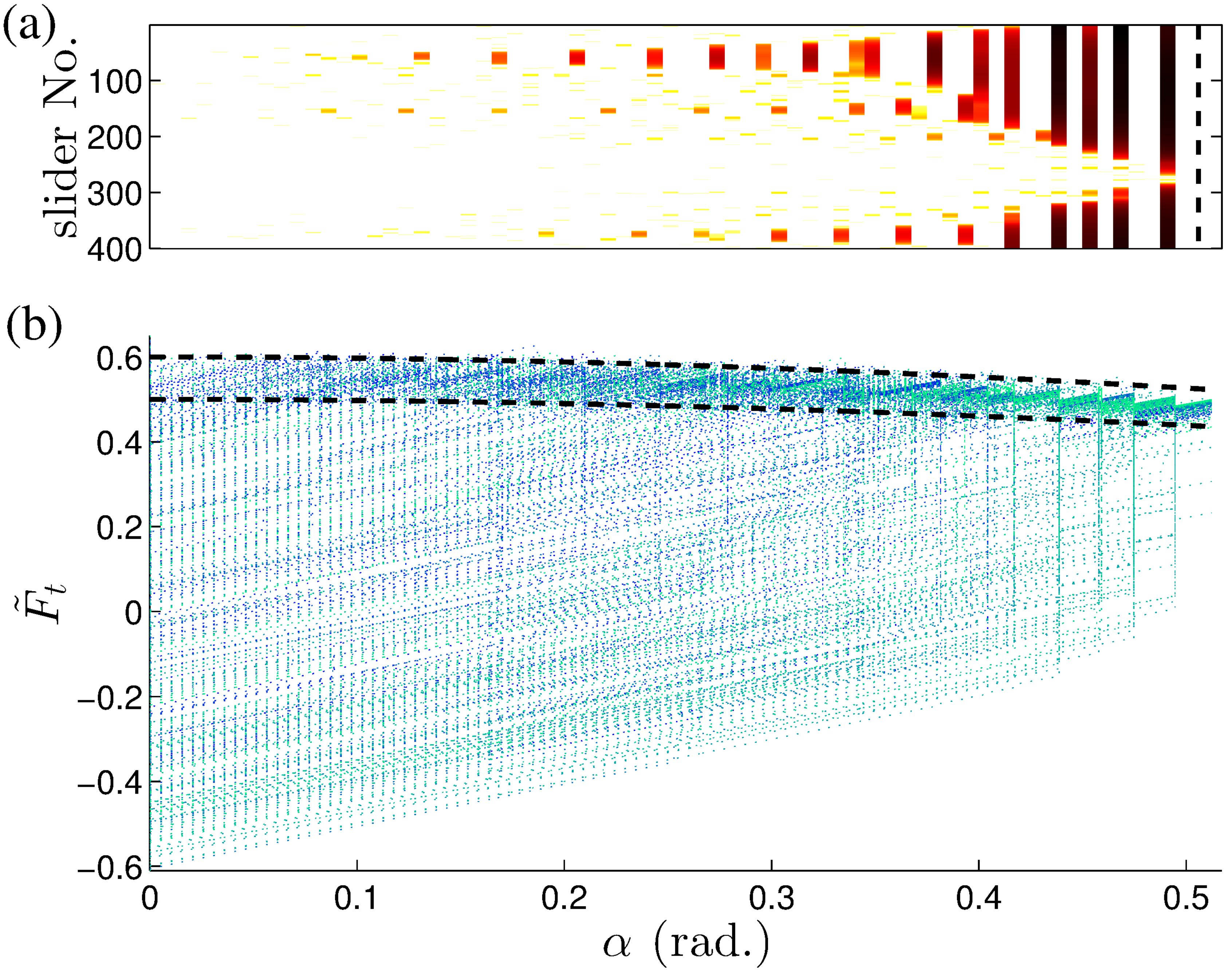}
  \caption{Behavior of a large system with global coupling upon
    inclination -- (a) Hot map of the displacements $\Delta \tilde
    x_i$ vs. angle $\alpha$. (b) Tangential force $\tilde F_t$
    vs. angle $\alpha$ (The slider number is indicated by the color
    bar).  The upper dashed line is the stability limit: $\bar{\mu}_s
    \cos \alpha$.  The lower dashed line is the arrest limit $(2\mu_d
    - \bar{\mu}_s) \cos \alpha$ [$N = 10$, $\xi = 0.001$, $l=10$,
      $\mu_d = 0.55$, $\bar{\mu}_s = 0.6$, and $\sigma_{\mu} =
      0.01$].}
  \label{fig:large_wc}
\end{figure}

Comparing the behaviour of the system with (Fig.~\ref{fig:large_wc})
and without (Fig.~\ref{fig:large_no}) coupling, we immediately notice
the coupling leads to larger rearrangments, involving more sliders and
larger displacements than in the uncoupled system.  In addition, in
Fig.~\ref{fig:large_wc}, we observe two main active regions involving
the two extremities of the system.  From an angle of about $\alpha
\sim 0.4$, the upper region extends over more than half the size of
the system.  We also note also that the two active regions synchronize
$\alpha \gtrsim 0.42$ even if they do not overlap until the avalanche
of the whole system which occurs for $\alpha \simeq 0.51$.

In Fig.~\ref{fig:large_wc}b, we observe that the tangential components
of the forces behave as previously observed in absence of coupling
(Fig.~\ref{fig:large_no}b).  Those components, for the sliders that
moved at least once, are bounded by two well-defined values.  The
effect of the global coupling is clearly visible in the evolution of
the tangential component of the forces for $\alpha \gtrsim 0.42$: for
each rearrangment, the tangential component corresponding to the
sliders that are not involved exhibit a large increase.  This
synchronization is also visible by focussing on the behavior of the
forces between the two dashed lines in Fig.~\ref{fig:large_no}b which
correspond to the sliders that moved.

\section{Analytical model}
\label{sec:model}

We present in this section a simple approach which allows to
understand the main features that have been observed in the
simulation. First we will study the case of a single slider surrounded
by two immobile blocks (Sec.~\ref{sec:model_one}).
In Sec.~\ref{sec:model_stat}, we will see how a
statistical model can be built from the results obtained with a single slider.
In this section no global coupling is ever considered ($\xi = 0$).
The effect of the coupling, as well as the comparison with experiments,
will be the subject of the discussion (Sec.~\ref{sec:discussion}).

\subsection{A single slider}
\label{sec:model_one}

\begin{figure}[t!]
  \centering \includegraphics[width=\columnwidth]{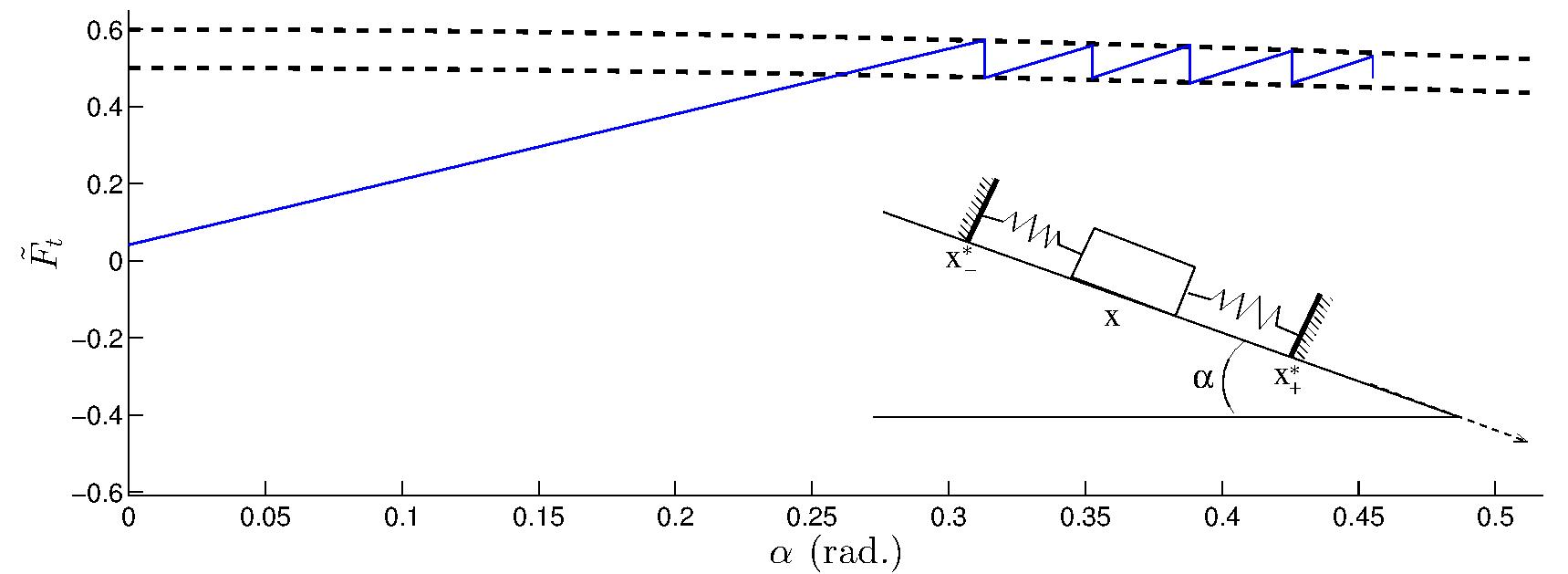}
  \caption{Tangential force $\tilde F_t$ vs. angle $\alpha$ -- The
    upper dashed line is the stability limit: $\bar{\mu}_s \cos
    \alpha$.  The lower dashed line is the arrest limit $(2\mu_d -
    \bar{\mu}_s) \cos \alpha$.  Inset: sketch of the system under
    study [$\mu_d = 0.55$ and $\mu_d = 0.6$].}
  \label{fig:schema_one}
\end{figure}

Let us first consider a single slider bonded by two springs to two immobile walls
as sketched in Fig.~\ref{fig:schema_one}.

We consider that the walls are located at $x_{-}^*$ and $x_{+}^*$.
Initially, the system is horizontal ($\alpha = 0$) and the slider at
stable position $x^{(0)}$. For the sake of simplicity, we assume that
static frictional coefficient is constant and equal to
$\mu_s$. Indeed, in practice $\sigma_\mu \ll 2\mu_s$, such that the
associated width of the distribution of the static frictional force
can be neglected when compared to the variation of the tangential
force associated to the choice of $x^{(0)}$.

From Eq.~\eqref{eq:stability}, we can determine the angle $\alpha^{(1)}$ corresponding
to the limit of stability of the slider:
\begin{align}
\tilde F_t^{(0)} + \sin \alpha^{(1)} = \mu_s &\cos \alpha^{(1)},\\
\mathrm{where~}\tilde F_t^{(0)} &= \tilde x_{+}^{*} + \tilde x_{-}^{*} - 2\tilde x^{(0)}\nonumber
\end{align}
Once $\alpha^{(1)}$ is reached by tilting the system, for this simple case
the dynamical Eqs.~\eqref{eq:dynamics} can be solved analytically, which 
leads to: 
\begin{equation}
\Delta \tilde x^{(1)} \equiv \tilde x^{(1)} - \tilde x^{(0)} = (\mu_s - \mu_d) \cos \alpha^{(1)}
\end{equation}
The total tangential force exerted on the block in its new position is
\begin{align}
\tilde F_t^{(1)} = \mu_s \cos \alpha^{(1)} &+ \Delta \tilde F^{(1)},\\
\mathrm{where~}\Delta \tilde F^{(1)} &= - 2 (\mu_s - \mu_d) \cos \alpha^{(1)}\nonumber
\end{align}

In the same way, upon further inclination, the slider destabilizes for $\alpha^{(2)}$ verifying:
\begin{equation}
\left( \tilde F_t^{(1)} - \sin \alpha^{(1)} \right) + \sin \alpha^{(2)} = \mu_s
\cos \alpha^{(2)}
\end{equation}
Assuming that $\Delta \alpha^{(2)} = \alpha^{(2)} - \alpha^{(1)}$ is small, we get to the first order:
\begin{equation}
\Delta \alpha^{(2)} = \frac{2 (\mu_s - \mu_d)}{1 + \mu_s \tan
  \alpha^{(1)}}
\end{equation}
which remains small provided that $\mu_s - \mu_d \ll 1$.
The reasoning can be reproduced for any of the rearrangments, leading to:
\begin{equation}
\Delta \alpha^{(i+1)} = \frac{2 (\mu_s - \mu_d)}{1 + \mu_s \tan \alpha^{(i)}}
\end{equation}

Thus, a generic behavior emerges as shown in Fig.~\ref{fig:schema_one}.
We observe, first, a loading phase, which depends on the initial position $\tilde x^{(0)}$.
The tangential component of the force $\tilde F_t$ increases continuously
until it reaches the instability threshold for $\alpha = \alpha^{(1)}$
and suddenly drops by $\Delta \tilde F_t^{(1)}$.
Then, rearrangements occur for successive
values $\alpha^{(i)}$ of the inclination angle. They are marked by
sudden drops $\Delta \tilde F_t^{(i)}$ of the tangential component,
followed by continuous increase towards the next threshold.
In a first order approximation in $\alpha$, the force drops $\Delta \tilde F^{(i)}$
do not depend on the initial position of the slider.
In addition, for small enough angle $\alpha^{(i)}$, the interval $\Delta \alpha^{(i+1)}$
is almost constant and equal to $\Delta \alpha \simeq 2\,(\mu_s - \mu_d)$.
Such periodic behaviour is reminiscent of the well-known stick-slip motion.
Finally, note that a sliding event leading to a drop $\Delta \tilde F_t^{(i)}$ of the force
exerted to the slider, results in an increase
by $\Delta \tilde F_t^{(i)}/2 = (\mu_s - \mu_d) \cos \alpha$
of the force exerted on each of the walls.

%%%%%%%%%%%%%%%%%%%%%%%%%%%%%%%%%%%%%%%%%%%%%%%%%%%%%%%%%%%%%%%%%%%%%%%%%%%%%%%%%%%%%%%%%%%%%%%%%%%%%%%%%%%%%%%%%%%%%%%%%%%%%%%%%%%%%%%%%%%%%%%%%%%%%%%%%%%%%%%%%%%%%%%%%
\subsection{Statistical point of view}
\label{sec:model_stat}

As will be seen in the discussion (Sec.~\ref{sec:discussion}),
the simple analysis of a single slider between two walls
helps to understand most of the features observed for a small system.
For large systems however, where the pattern that
emerges is more complicated (Sec.~\ref{sec:results_large}),
a statistical approach is more appropriate to model the system.

In the present section, we seek for master equations describing the
evolution of the distribution of the tangential forces $\tilde F_t$ in
a system as a function of the inclination angle $\alpha$.  We will
make strong approximations to be able to provide an analytical model,
but we will compare the solutions we obtain to numerical simulations
of large systems in absence of global coupling ($\xi=0$).

\subsubsection{Qualitative analysis}
\label{sec:qualitative}

\begin{figure}[htbp]
  \centering
  \includegraphics[width=\columnwidth]{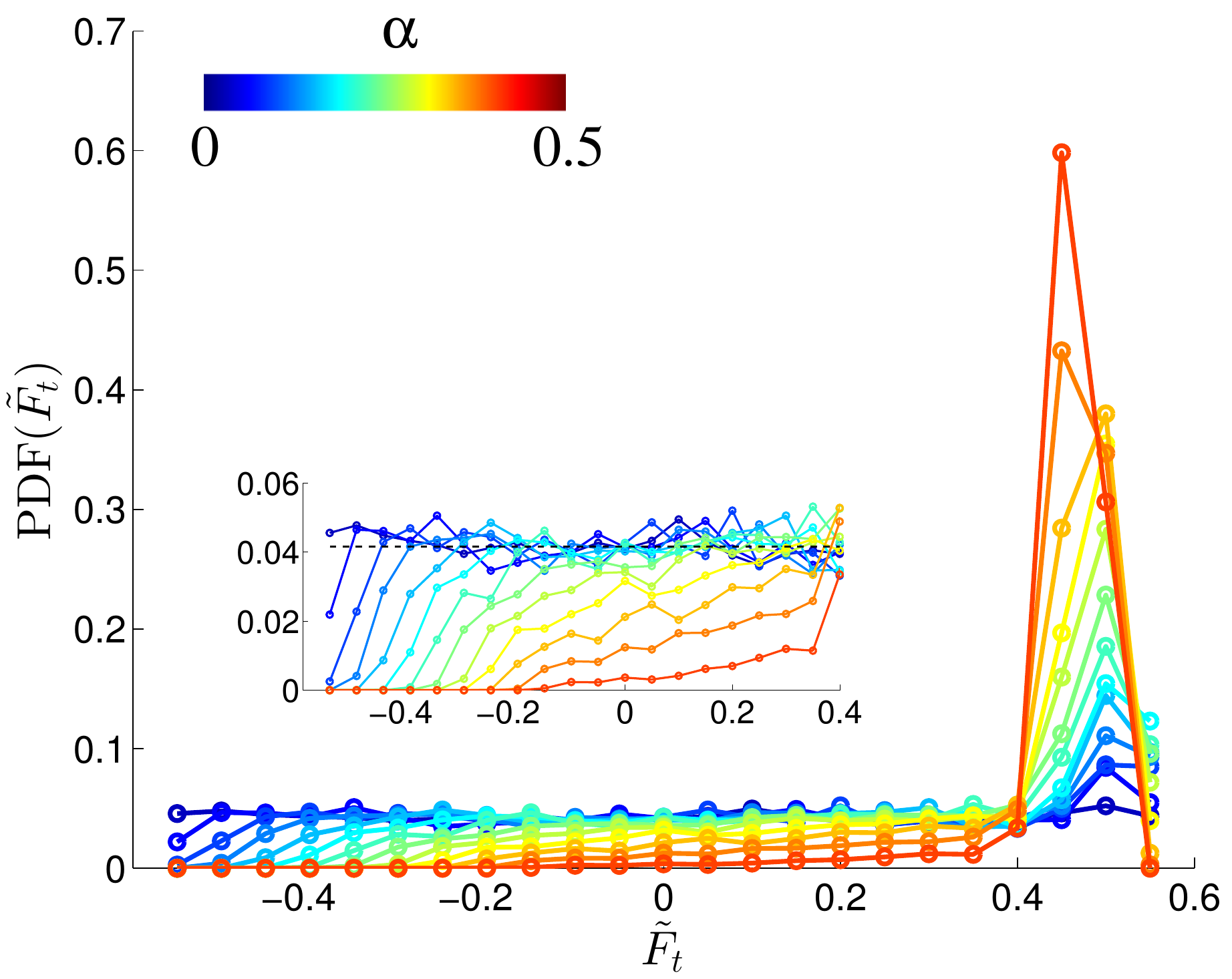}
  \caption{Probability distribution function of the tangential force
    $\tilde F_t$ upon inclination -- at different values of the
    inclination angle -- Each distribution corresponds to an average,
    at a given $\alpha$ (indicated by the colorscale), over 20
    numerical runs.  Inset: Enlargment of the distribution for $\tilde
    F_t < 0.4$ [$N = 100$, $\xi = 0$, $\ti{l}=10$, $\mu_d = 0.55$,
      $\bar{\mu}_s = 0.6$, and $\sigma_{\mu} = 0.01$].}
  \label{fig:distrib1}
\end{figure}

We report in Fig.~\ref{fig:distrib1} the probability distribution function
of the tangential forces, $\tilde F_t$, exerted on the sliders at different values of
the inclination angle $\alpha$, in absence of global coupling ($\xi =0$).

Initially, at $\alpha=0$ (dark blue curve in Fig.~\ref{fig:distrib1}),
the distribution is flat in the interval $]-\mu_s,\mu_s[$, which
corresponds to the initial, random, preparation of the system at horizontal
(Sec.~\ref{sec:method_3}).

Upon inclination, we observe that the distribution functions exhibits two main parts:
\begin{list}{-}{}
 \item A rather flat part for $\tilde F_t \lesssim 0.4$ with a well-defined plateau at
 the constant initial value. The lowest values of $\tilde F_t$ are slowly depleted.
 \item A peak, for $0.4 \lesssim \tilde F_t \lesssim \mm$, that grows rapidly.
\end{list}
The qualitative behaviour of the distribution function can be
understood as follows. Upon inclination, the lowest values of $\tilde
F_t$ are depleted because of the increase of the projection of the
weight along the incline. By this simple effect, the lower bound of
the distribution increases according to $-\mm \cos \alpha + \sin
\alpha$.  Conversely, upon increasing inclination, sliders loose
stability.  The upper bound, which equals $\mm \cos \alpha$ according
to (Eq.~\ref{eq:stability}), slightly decreases, but this variation
can be neglected at the first order.  The value of $\tilde F_t$ for
the sliders that moved remains close to the upper bound. The peak
grows as their number increases.  This picture holds as long as the
plateau and the peak are well separated, i.e. for $\alpha \lesssim
0.25$. Then, the contribution of the sliders that are involved in the
rearrangements become predominent.

\begin{figure}[htbp]
  \centering
  \includegraphics[width=\columnwidth]{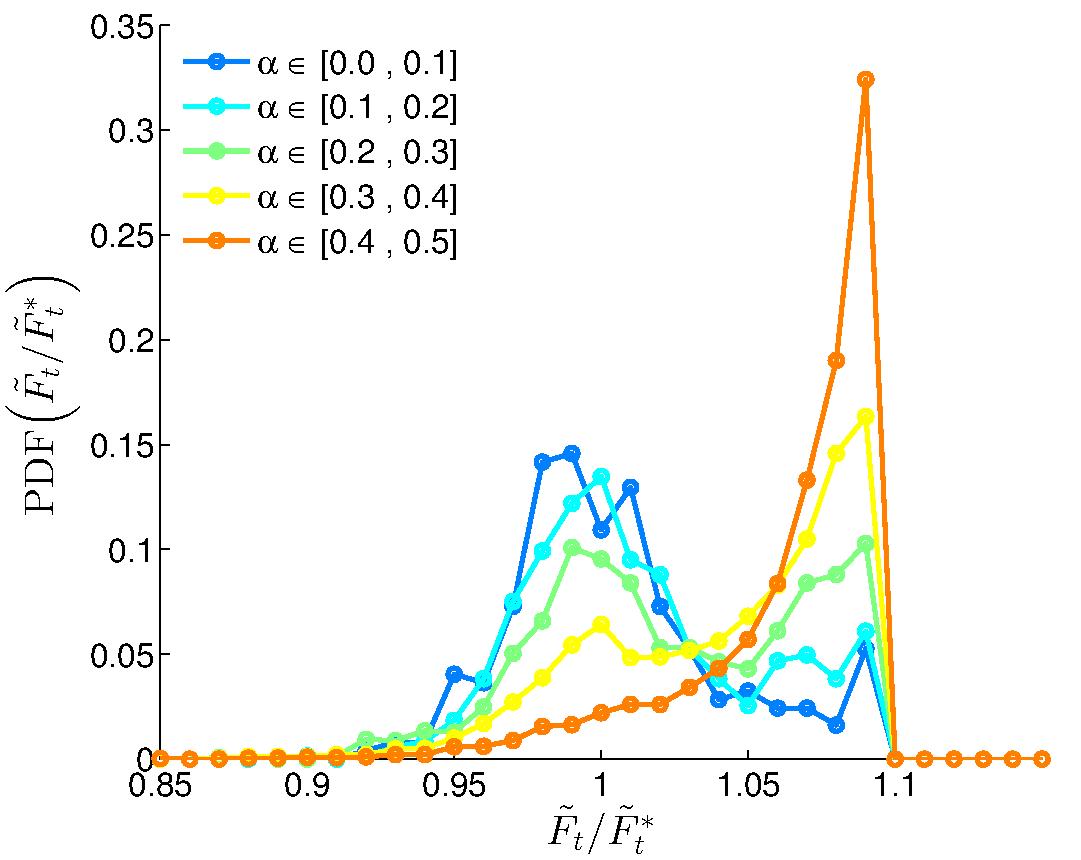}
  \caption{Probability distribution function of the ratio $\tilde
    F_t/\tilde F_t^*$ for the sliders that moved upon inclination --
    The data are from the 20 same numerical runs as in
    Fig.~\ref{fig:distrib1} [$N = 100$, $\xi = 0$, $\ti{l}=10$, $\mu_d
      = 0.55$, $\bar{\mu}_s = 0.6$, and $\sigma_{\mu} = 0.01$].}
  \label{fig:distrib3}
\end{figure}

%%%%%%%%%%%%%%%%%%%%%%%%%%%%%%%%%%%%%%%%%%%%%%%%%%%%%%%%%%%%%%%%%%%%%%%%%%%
In order to get more insights in the contribution of the
rearrangments, we focus on the distribution of the values of $\tilde
F_t$ for only the sliders that moved (Figure~\ref{fig:distrib3}).
Defining $\tilde F_t^* \equiv (2\mu_d - \mm) \cos \alpha$, we observe
that, for small inclination, the distribution function is peaked
around $\tilde F_t/\tilde F_t^* =1$.  This observation shows that the
rearrangements initially involve single sliders, the width of the peak
is directly linked to the width of the distribution of $\mu_s$. By
contrast, for large inclination, the distribution function exhibits a
peak close to the value $\tilde F_t/\tilde F_t^* \simeq 1.1$, which is
an upper limit for all the distribution functions.

We can understand the accumulation of the tangential forces at this
limit value in the following way.  First, let us mention that $\tilde
F_t/\tilde F_t^* \simeq 1.1$ corresponds to $\tilde F_t = \mu_d \cos
\alpha$.  For rearrangements implying several interacting sliders, the
movement becomes complex enough for the vanishing velocities not to be
determined by a sinusoidal movement, but by the overall slowing down
of the dynamics due to dissipation. At the limit of vanishing
accelerations, the sum of elastic forces and weight on each block
compensates the dynamical friction so that, when the block stops, the
value of the resultant tangential force is typically less than $\mu_d
\cos \alpha$, thus controlled by the dynamical frictional coefficient.

In conclusion, after a sliding event, for the sliders implied in the event,
the tangential forces $\tilde F_t$ drop by a quantity comprised between
$\Delta \mu \cos \alpha$ and $2\Delta \mu \cos \alpha$ with $\Delta \mu = \mm - \mu_d$.
This is the main ingredient of the model we present in the next section \ref{subsec:model_stat}.

%%%%%%%%%%%%%%%%%%%%%%%%%%%%%%%%%%%%%%%%%%%%%
\subsubsection{Model}
\label{subsec:model_stat}

\begin{figure}[htbp]
  \centering
  \includegraphics[width=\columnwidth]{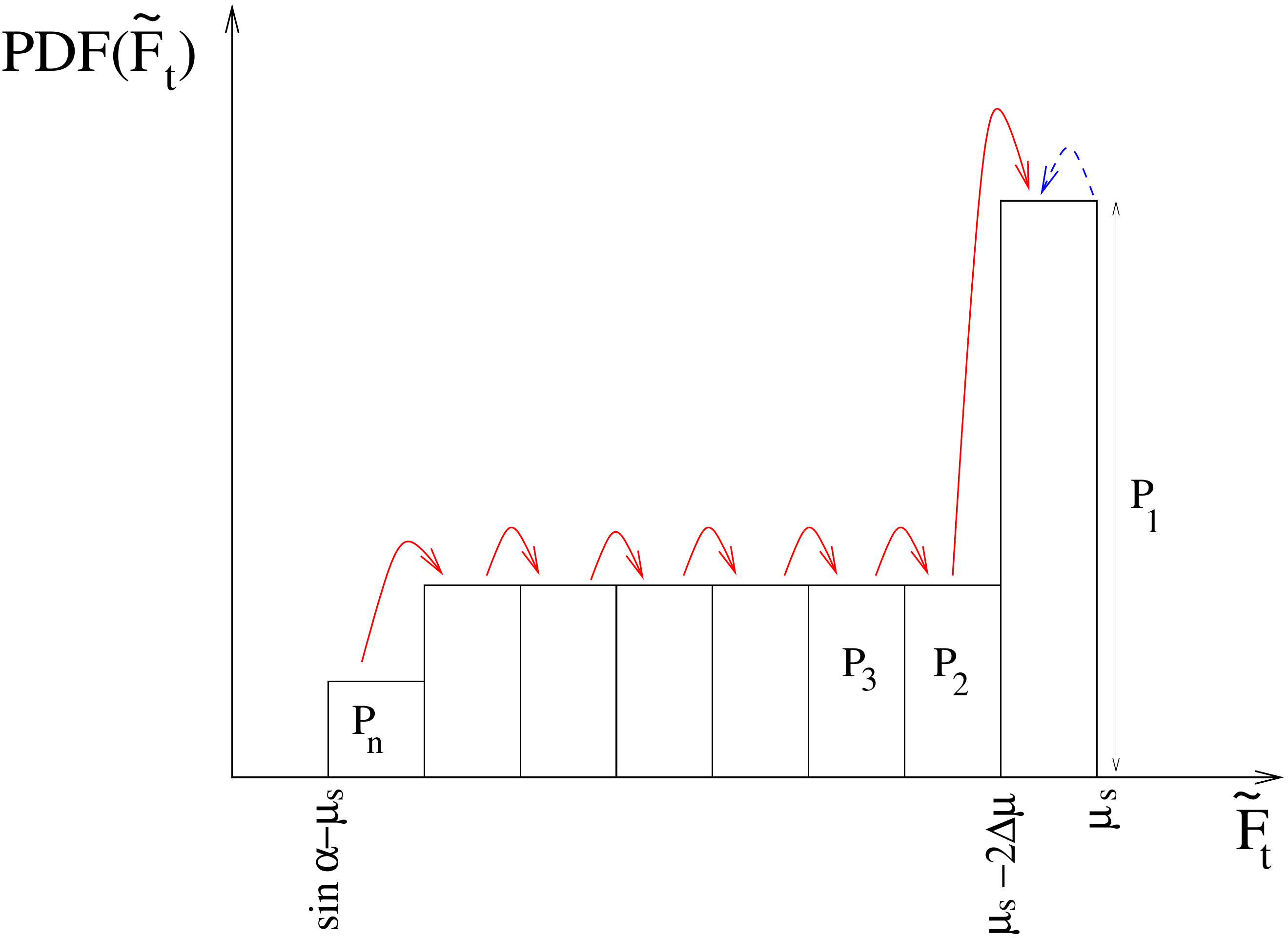}
  \caption{Schematic representation of the histogram equivalent to
    the probability distribution function of the tangential force,
    $\tilde F_t$ at a given inclination angle $\alpha$.  The arrows
    indicate the main modification of the distribution upon further
    inclination by $\delta \alpha$ (see
    Sec.~\ref{subsec:model_stat}).}
  \label{fig:sch_d}
\end{figure}

Here we present a model based on the prior observations.  We consider
in Fig.~\ref{fig:sch_d} a schematic representation as a bar histogram
of the distribution function reported in Fig.~\ref{fig:distrib1}.  To
do so, we discretize the probability density function in bins of width
$2\Delta \mu$ with $\Delta \mu = \mm - \mu_d$.  In addition, in a
first approach, we neglect the effect of the inclination on the upper
bound, which we assume to equal $\mm$.  We denote $\nu$ the number of
bins of width $2\Delta\mu$, starting from the upper bound, $\mm$, and
including the lower bound, $\sin \alpha - \mm$.  Thus, $\nu$ is a
function of $\alpha$ that decreases with $\alpha$ as: $\nu(\alpha) =
\lceil{\frac{2\mm - \sin \alpha}{2 \Delta \mu}}\rceil$. For
convenience, the bins are numbered from large to small forces such
that the bin No. 1 denotes the peak on the right-hand-side,
corresponding to the peak of the distribution.  We denote $P_i$ the
height of bin No. i.  With these definitions, the probability for a
slider to be submitted to a tangential force belonging to the bin No. i
is given by $2\Delta \mu P_i$.

As a result of an angular increment $\delta \alpha$,
all the tangential forces increase because of the increase
of the tangential component of the weight which goes like $\sin\alpha$.
Consequently, in the distribution, $N \cos \alpha P_i \delta \alpha$
sliders shift from the bin $i$ to the bin $i-1$, excepted for the bin No. 1.
This process is sketched by the red solid arrows in Fig.~\ref{fig:sch_d}.
For the bin No. 1, $N\cos\alpha P_1 \delta \alpha$ sliders loose stability.
From the qualitative description of the system (Sec.~\ref{sec:qualitative}),
we know that the final value of $\tilde F_t$ after the rearrangement
lies mostly between $\mu_d \cos \alpha$ and $(\mu_s - 2\Delta \mu) \cos \alpha$.
Thus, after the rearrangments, the sliders remain in bin No. 1.
This process is sketched by the blue dashed arrow in Fig.~\ref{fig:sch_d}.

This simple process can be put in equations in the form:
\begin{align}
\frac{dP_{\nu}}{d\alpha} &= -\frac{1}{2\Delta \mu}P_{\nu}\nonumber\\
\frac{dP_i}{d\alpha} & = \frac{1}{2\Delta \mu} (P_{i+1} - P_i)~~~~~(i\neq1,\nu)\\
\frac{dP_1}{d\alpha} & = \frac{1}{2\Delta \mu} P_2\nonumber
\end{align}
Starting from the condition that the distribution is uniform for $\alpha = 0$,
thus from $P_i(0) = \frac{1}{2\mm}$, we get:
\begin{align}
P_{\nu} (\alpha) & = \frac{1}{2\mm} e^{-\frac{\alpha}{2\Delta \mu}}
\nonumber \\
P_i(\alpha) & = \frac{1}{2\mm} \left[
  \sum_{k=0}^{\nu - i} \frac{1}{k!} \left(\frac{\alpha}{2 \Delta
    \mu}\right)^k \right] e^{-\frac{\alpha}{2\Delta
    \mu}} \label{eq:model}\\
P_1(\alpha) & = 
\frac{1}{2\mm} \left\{ \nu - 1 - \sum_{k=0}^{\nu - 2} \left[ \sum_{j=0}^k \frac{1}{j!}
  \left(\frac{\alpha}{2 \Delta \mu}\right)^j \right]
e^{-\frac{\alpha}{2\Delta \mu}} \right\} \nonumber
\end{align}

It is of particular interest to focus on the behavior of the solution
far from the lower boundary.  For realistic values of $\mu_s$ and
$\mu_d$, $\nu(\alpha)$ remains large and, in the limit, $\nu - i \gg
1$, we have $\sum_{k=0}^{\nu - i} \frac{x^k}{k!}  \simeq
e^x$~($\forall x$). Consequently, for small angle $\alpha$ so that
$\nu(\alpha)$ is still large ($\nu(0) = \frac{\mm}{\Delta \mu} \gg 1$
for realistic values of $\mu_s$ and $\mu_d$) and for $i\ll \nu$,
i.e. far from the lower bound of the system, we have:
\begin{equation}
 \begin{aligned}
P_i(\alpha) & \simeq \frac{1}{2\mm}~~~(i \neq 1) \land (i \ll \nu)\\
P_1(\alpha) & \simeq \frac{1}{2\mm} \left(1+\frac{\alpha}{2\Delta \mu}\right)
\label{eq:crude}
\end{aligned}
\end{equation}
Note that, as is, $P_1(\alpha)$ accounts for the amplitude of the peak
at large $\tilde F_t$ as function of the inclination
whereas the constant $P_i(\alpha)$ correspond to the plateau.

\begin{figure}[htbp]
  \centering
  \includegraphics[width=\columnwidth]{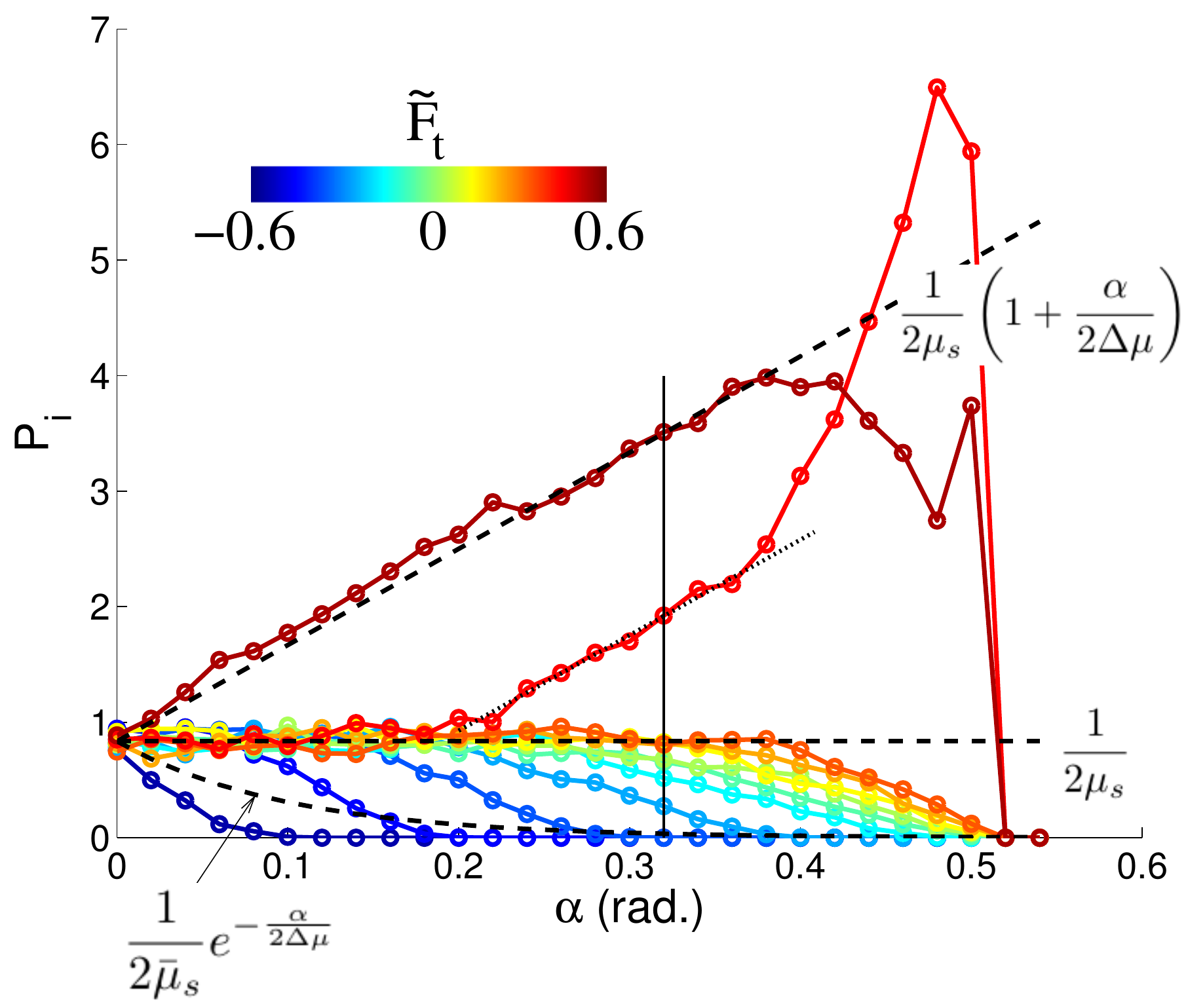}
  \caption{Height $P_i$ vs. angle $\alpha$ for the data of
    Fig.~\ref{fig:distrib1}.  The dashed straight lines correspond to
    prediction of the crude model (Eq.~\ref{eq:crude}).  The
    exponential decay of $P_\nu(\alpha)$ is given in
    Eq.~(\ref{eq:model}).  The dotted line is a guide for the eye.
    The vertical solid line is the threshold from Eq.~\eqref{eq:cond}
    [$N = 100$, $\xi = 0$, $\ti{l}=10$, $\mu_d = 0.55$, $\bar{\mu}_s =
      0.6$, and $\sigma_{\mu} = 0.01$].}
  \label{fig:distrib2}
\end{figure}

We can now compare our model (Eq.~\ref{eq:model}) and its limiting case (Eq.~\ref{eq:crude}) to the results
of the numerical simulations. To do so, we report in Fig.~\ref{fig:distrib2},
the probability $P_i$ as function of the inclination angle $\alpha$ for the data already
reported in Fig.~\ref{fig:distrib1}.

First, we observe that Eq.~(\ref{eq:crude}) correctly describes the
linear increase of the amplitude of the peak $P_1$ at large $\tilde
F_t$ in a large range of inclination angle $\alpha$ (0 to typically
$0.38$).  Second, Eq.~(\ref{eq:crude}) also accounts for the constant
value of $P_i$ in the same range of inclination, but only for the
largest values of $\tilde F_t$.  Slightly less crude,
Eq.~(\ref{eq:model}) predicts the decrease of the $P_i$ ($i \neq 1$)
upon increasing $\alpha$. However, reporting the probability of the
smallest value of $\tilde F_t$, we observe the model underestimates
the depletion of the smallest values of $\tilde F_t$. This point will
be further commented in the discussion (Sec.~\ref{sec:discussion}).

In spite of this reasonably good agreement between the model and the numerical
results, a trend observed in Fig.~\ref{fig:distrib2}, which is not described
at all by the model, is the linear increase of $P_2$ starting at $\alpha \simeq 0.2$,
followed by a sharp increase for $\alpha \gtrsim 0.4$.
We observe that, for $0.2 \lesssim \alpha \lesssim 0.4$,
the slope $dP_2/d\alpha = \frac{1}{4\mu_s \Delta \mu}$, thus takes the same value that $dP_1/d\alpha$.
 The model fails in accounting for the increase of $P_2$ because we did not consider
the decrease of the upper bound upon inclination (the maximum force value evolves like $\cos \alpha$).

As a matter of fact, due to the shift of the upper bound, the bin
No. 2 takes the place of bin No. 1 as the inclination proceeds.

Up-to-now, we commented mainly on the behavior of the system for limited
values of the inclination angle $\alpha \lesssim 0.38$.
For $\alpha \gtrsim 0.38$, a drastic change in the behavior
of all the probabilies $P_i$ is observed.
Indeed, the probability of the largest values, i.e. $P_1 + P_2$, drastically increases
whereas the probabilities of smaller values $P_i$ ($i > 2$) decreases.
One interesting question is whether our model is able to predict the typical
angle, $\alpha_t \simeq 0.4$, at the transition.
To answer the question, consider the probability $P_3$ of the force just below the peak:
\begin{align*}
P_3(\alpha) & = \frac{1}{2\mu_s} \left[ \sum_{k=0}^{\nu(\alpha) - 3}
  \frac{1}{k!} \left(\frac{\alpha}{2 \Delta \mu}\right)^k \right]
e^{-\frac{\alpha}{2\Delta \mu}}\\
 & \simeq \frac{1}{2\mu_s} \left( 1-
\frac{1}{(\nu(\alpha)-2)!} \left(\frac{\alpha}{2\Delta \mu}
\right)^{\nu(\alpha)-2}\right)
\end{align*}
We previously considered that $\nu(\alpha)$ was large enough for the term
on the right-hand-side in the parenthesis to be negligible. Let assume that
this is not the case anymore and that this term is of 1/10, i.e. that:
\begin{equation}
\frac{1}{(\nu(\alpha)-2)!} \left(\frac{\alpha}{2\Delta \mu}
\right)^{\nu(\alpha)-2} = 0.1. \label{eq:not_neg}
\end{equation}
Using the Stirling approximation and the definition of $\nu(\alpha)$,
we get:
\begin{equation}
\alpha_t \simeq \frac{2 \mu_s}{1 + e} \simeq 0.54\,\mu_s
\label{eq:cond}
\end{equation}
which, we point out, does not significantly depend on the choice
of the value $1/10$ in Eq.~\eqref{eq:not_neg}.
We thus get $\alpha_t \sim 0.32$ for $\mm = 0.6$, 
which underestimates the observed value.
However, in regard of the crude approximations we made, we can consider
that this last estimate is reasonable.

%%%%%%%%%%%%%%%%%%%%%%%%%%%%%%%%%%%%%%%%%%%%%%%%%%%%%%%%%%%%%%%%%%%%%%%%
%%%%%%%%%%%%%%%%%%%%%%%%%%%%%%%%%%%%%%%%%%%%%%%%%%%%%%%%%%%%%%%%%%%%%%%%
%%%%%%%%%%%%%%%%%%%%%%%%%%%%%%%%%%%%%%%%%%%%%%%%%%%%%%%%%%%%%%%%%%%%%%%%
\section{Discussion}
\label{sec:discussion}

\subsection{Discussion of the model}

The very simple model of a single slider (Sec.~\ref{sec:model_one})
provides a good understanding of the occurence of the, local,
quasi-periodic dynamics since the beginning of the inclination
process, i.e. for small tilt angle $\alpha$ (typically~$\leq 0.2$).
Even in a large system, the first sliders that destabilize are
isolated from one another and behave like the single slider between
two walls, which consist of the two immobile neighbors.  The
tangential force applied to it increases until it reaches the
threshold $\mm \cos \alpha$ [upper dashed line in
  Fig~\ref{fig:small}(b)] and then drops by $2(\mm - \mu_d) \cos
\alpha$, then reaching the value $(2\mu_d - \bar{\mu}_s) \cos \alpha$
      [lower dashed line in Fig~\ref{fig:small}(b)]. Each rearrangment
      leads to the increase of the tangential force applied to the
      neighbors by $(\mm - \mu_d) \cos \alpha$.

Upon inclination, rearrangments involving initially only one slider
lead to an increase of the tangential force applied to its neighbors
sufficient to destabilize at least one of them.  The pattern is then
more complex but the picture of a process consisting of a series of,
quasi-periodic, localized destabilizations holds.  The number of
sliders involved in each rearrangment increases upon inclination. When
several sliders are involved, the rearrangements repeat with the
period $\Delta \alpha \sim (\mm - \mu_d)$ rather than $\Delta \alpha
\sim 2(\mm - \mu_d)$ (Fig~\ref{fig:small}).  Indeed, as discussed in
Sec.~\ref{sec:qualitative}, when several sliders enter in motion, the
progressive slowing down of the dynamics due to dissipation leads to
the value $\mu_d \cos \alpha$ [instead of $(2\mu_d - \bar{\mu}_s) \cos
  \alpha$ for a single slider] of the tangential force.

However, in large systems, in absence of global coupling,
the previous picture holds only locally
and the overall dynamics is not quasi-periodic (Fig.~\ref{fig:large_no}). 
The rearrangments occuring in different regions are not correlated.
The growth in the size of the active regions upon inclination leads
them to merge before the whole system looses stability.

The introduction of a global coupling, even small ($\xi = 10^{-3}$),
leads to the synchronisation of the previously isolated local rearrangements
and, thus, to the occurence of a quasi-periodic dynamics of the whole system (Fig.~\ref{fig:large_wc}).
This synchronization is reminiscent of the synchronization of oscillators, which is a well-known
phenomenon~\cite{Strogatz1993}.

A particularly interesting, and striking, result is that one can
distinguish a clear change in the activity of the system, with or
without global coupling, at an inclination angle $\alpha$ of about
$0.25$, thus typically half of the avalanche angle $\alpha_c$: for
$\alpha \lesssim 0.25$, the system is rather quiescent with few
rearrangements occuring locally whereas, for larger $\alpha$, numerous
large events of increasing size occur.  The model shows that a change
in the dynamics is indeed expected for $\alpha_t \simeq 0.54\,\mm$.
For $0 \leq \alpha \lesssim \alpha_t$, the number of active sliders
increases linearly with $\alpha$. By contrast, for $\alpha \gtrsim
\alpha_t$, the number of active sliders drastically increases
(Fig.~\ref{fig:distrib2}).  In practice, $0.54 \mm$ is close to half
the angle of avalanche angle, $\alpha_c$, so that we can consider
that, to the level of approximation of our model, the transition is
well accounted for.

\subsection{Comparison to experiments}

Finally, we underline strong similarities between features of our
model and the experiments mentionned in the
introduction~\cite{Nerone2003,Kiesgen2012,Amon2013}.

Upon inclination of a container filled with a granular material,
one observed, for small tilt angle, the occurence of small, uncorrelated,
rearrangements randomly distributed in the system. At approximately
half the avalanche angle, the activity drastically increases
in the form of bursts of synchronized displacements. 

The order of magnitude of the angular period observed experimentally
is typically of a few degrees. In our model, we predict that the
periodicity is linked to the difference between the static and the
dynamic friction coefficients: $\Delta \alpha \sim (\mu_s - \mu_d)$.
For granular matter, those friction coefficients can be assimilated to
the tangent of, respectively, the angle of avalanche, at which the 
granular surface looses stability, and of the angle of repose,
at which corresponds the granular surface stops flowing.
In granular matter $(\mu_s - \mu_d) \sim 0.1$~\cite{Jaeger1989}
leading thus to $\delta \alpha$ indeed of a few degrees.

In our model, we obtain synchronization only when some global coupling
is introduced, which leads to the question of the physical
significance of this global coupling in experiments. In the classical
Burridge-Knopoff model a rearrangement affects only the immediate
neighbors. Such local model is irrealistic for modeling faults as the
bulk materials at each side of the fault are of finite stiffness and
mediate long-range elastic forces. Taking into account such long-range
elastic redistribution of the stress after a reaarrengement has been
taken into account in Burridge-Knopoff models by coupling blocks
elastically to a various number of distant
neighbors~\cite{Xia2005}. Such long-range elastic coupling has also
been discussed in tribology-related studies (see~\cite{Amundsen2012}
and references therein). In granular matter, it has been shown that
when a local, plastic rearrangement takes place, an elastic response
with long-ranged redistribution of the stress is
observed~\cite{LeBouil2014b,McNamara2016}.  This justifies the
introduction of a long-range coupling in the model. Another clue is
provided by the observation, in experiments, that the pseudo-period of
the bursts depends on the cohesion in the
system~\cite{Amon2013,Duranteau2013}.  This is in accordance with a
picture of a synchronization due to a long-range coupling.

\begin{figure}[htbp]
  \centering
  \includegraphics[width=\columnwidth]{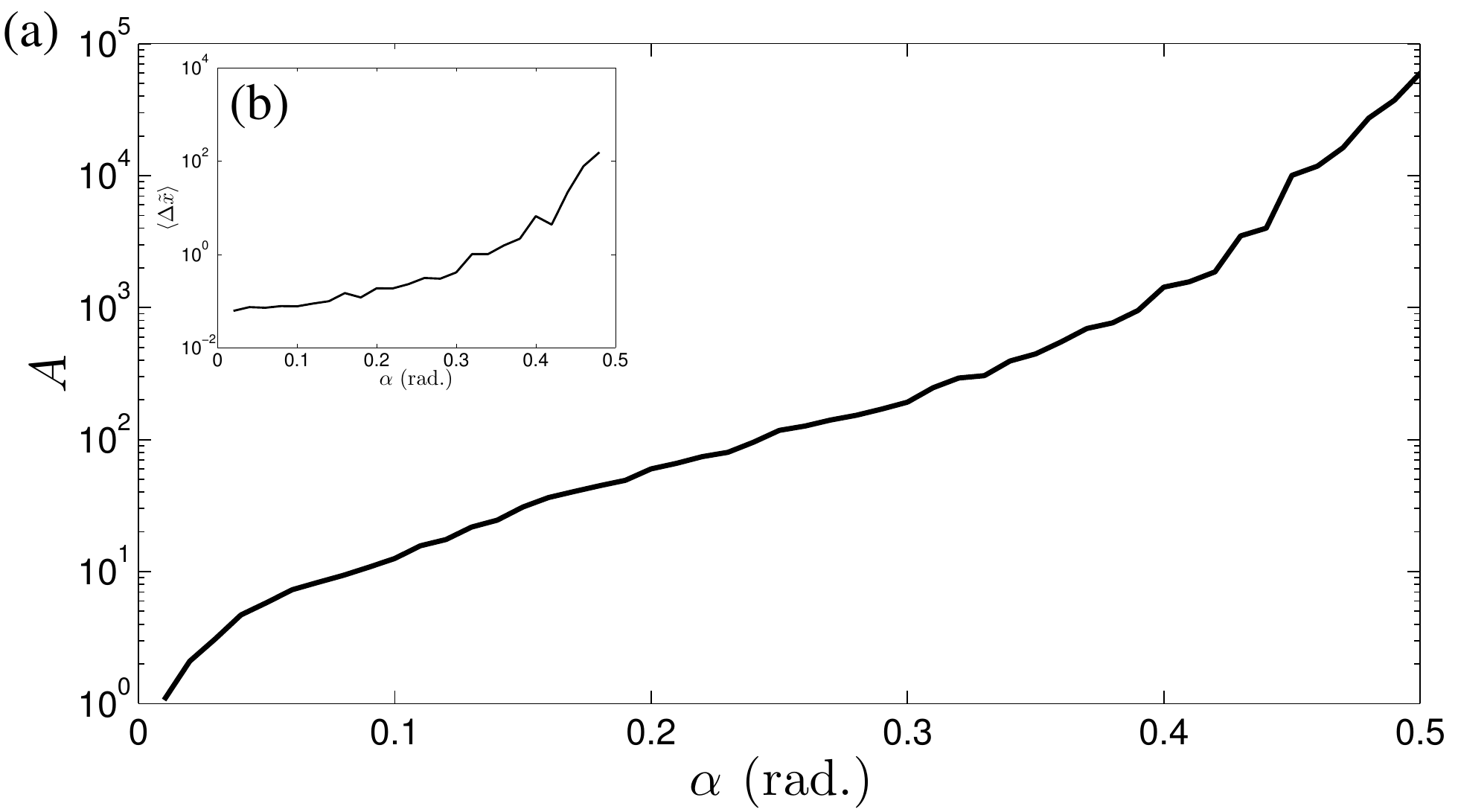}
  \caption{(a) Cumulated displacement $A$ vs. angle $\alpha$ -- Upon
    inclination, $A$ increases exponentially in spite of the linear
    increase of the number of sliders involved in the rearrangments
    with $\alpha$. (b) Average displacement during a rearrangement
    $\langle \Delta \tilde{x} \rangle$ vs. angle $\alpha$. [$N = 100$,
      $\xi = 0$, $\ti{l}=10$, $\mu_d = 0.55$, $\bar{\mu}_s = 0.6$, and
      $\sigma_{\mu} = 0.01$].} \label{fig:cumul}
\end{figure}

It has to be underlined that our model is not in contradiction with
previous works, in particular with studies based on cellular automata
reported in Ref.~\cite{Kiesgen2012}.  The authors reported an
exponential growth of the accumulated plastic activity upon
inclination.  This trend, also observed experimentally, has been
considered as a test validating the model.  Such an exponential
evolution could seem, at first sight, to be in contradiction with the
linear evolution of the number of active sliders which arises in our
model.  However, the contradiction is only apparent. Actually,
  reporting the cumulated displacements $A(\alpha)$ in the whole
  system from the beginning of the inclination process:
$$A(\alpha) = \sum_{i=0}^{i_c} \sum_{n=1}^N \left[
    \tilde{x}_n^{(i+1)}- \tilde{x}_n^{(i)} \right],$$ with
  $\tilde{x}_n^{(k)}$ the position of the $n^{th}$ slider after the
  $i^{th}$ rearrangement and $i_c$ the total number of rearrangements
  we also observe a typically exponential increase for our model
  (Fig.~\ref{fig:cumul}a). This is explained by an exponential
  increase of the typical displacement associated to the
  rearrangements with the inclination angle $\alpha$
  (Fig.~\ref{fig:cumul}b).

\section{Conclusion}

We have presented a one-dimensional frictional model
consisting of a chain of elastically coupled sliders
in frictional contact with an incline. 
The model reproduces most of the features observed experimentally
when quasi-statically tilting a box filled with a granular material.
In particular, the system reproduces the quasi-periodic
series of rearrangments observed in experiments.

We explain the regularity of the bursts of rearrangments
as the result of a combination of stick-slip and synchronization
due to a large scale coupling.
The phenomenon is thus related
to the problem of synchronization of oscillators.
The coupling originates from the redistribution of the forces,
at large scale, in the bulk of the material.
 
In addition, our toy model provides a statistical approach
to describe the evolution of the state of the system which approaches the avalanche.
In particular, we revealed an internal threshold before the avalanche occurs.
This threshold delimits the transition, at typically half the avalanche angle,
from an initial smooth increase
of the number of active sliders to an accelerating regime
where a dramatic increase of the activity takes place.
In regard to the prediction of avalanches, the determination of such
a threshold is precious as it delimits a rather regular and
predictible evolution of the system and a rapid growth regime
which announces the avalanche.

Our model remains simplistic and, in accordance, provides no more
than a qualitative description of the phenomenon.
In addition, it has to be noted that we considered a unique preparation,
consisting of a uniform distribution of the initial tangential forces.
Nevertheless, considering that the model is based on very general
arguments (disorder, solid friction and coupling),
we expect that it unveils the core mechanisms at play in different
experimental configurations.

\bibliographystyle{unsrt}

\begin{thebibliography}{}

\end{thebibliography}


\begin{thebibliography}{10}

\bibitem{Bak1987} P.~Bak, C.~Tang, and K.~Wiesenfeld, {\em
  Phys. Rev. Lett.} {\bf 59}, 381 (1987).

\bibitem{Nerone2003} N.~Nerone, M.~A. Aguirre, A.~Calvo, D.~Bideau,
  and I.~Ippolito, {\em Phys. Rev. E} {\bf 67}, 011302 (2003).

\bibitem{Kiesgen2012} S.~Kiesgen de~Richter, G.~Le Ca{\"e}r, and
  R.~Delannay, {\em J. Stat. Mech.: Theory Exp.}, (2012) P04013.

\bibitem{Amon2013} A.~Amon, R.~Bertoni, and J.~Crassous, {\em
  Phys. Rev. E} {\bf 87}, 012204 (2013).

\bibitem{Duranteau2013} M.~Duranteau, V.~Tournat, V.~Zaitsev,
  R.~Delannay, and P.~Richard, {\em AIP Conference Proceedings} {\bf 1542},
  650 (2013).

\bibitem{Gravish2014} N.~Gravish, and D.~~I. Goldman, {\em
  Phys. Rev. E} {\bf 90}, 032202 (2014).

\bibitem{Staron2002} L.~Staron, J.-P. Vilotte, and F.~Radjai, {\em
  Phys. Rev. Lett.} {\bf 89}, 204302 (2002).

\bibitem{Staron2006} L.~Staron, F.~Radjai, and J.-P. Vilotte, {\em
  J. Stat. Mech.} (2006) P07014.

\bibitem{Welker2011} P.~Welker and S.~McNamara, {\em Granular Matter}
  {\bf 13}, 93 (2011).

\bibitem{Nguyen2011} V.~B. Nguyen, T.~Darnige, A.~Bruand, and
  E.~Clement, {\em Phys. Rev. Lett.} {\bf 107}, 138303 (2011).

\bibitem{LeBouil2014a} A.~Le~Bouil, A.~Amon, J.-C. Sangleboeuf,
  H.~Orain, P.~B{\'e}suelle, G.~Viggiani, P.~Chasle, and J.~Crassous,
  {\em Granular Matter}, {\bf 16}, 1 (2014).

\bibitem{Rubinstein2007} S.~M. Rubinstein, G.~Cohen, and J.~Fineberg,
  {\em Phys. Rev. Lett.} {\bf 98}, 226103 (2007).

\bibitem{Burridge1967} R. Burridge and L. Knopoff, {\em
  Bull. Seismol. Soc. Am.} {\bf 57}, 3411 (1967).

\bibitem{Carlson1989} J. M. Carlson and J.S. Langer, {\em
  Phys. Rev. Lett.} {\bf 62}, 2632 (1989).

\bibitem{Carlson1994} J. M. Carlson, J.S. Langer, and B. E. Shaw, {\em
  Rev. Mod. Phys.} {\bf 66}, 657 (1994).

\bibitem{Braun2009} O. M. Braun, I. Barel, and M. Urbakh, {\em
  Phys. Rev. Lett.} {\bf 103}, 194301 (2009).

\bibitem{Maegawa2010} S. Maegawa, A. Suzuki, and K. Nakano, {\em
  Tribol. Lett.} {\bf 38}, 313 (2010).

\bibitem{Blanc2011} B.~Blanc, L.~A.~Pugnaloni, and J.-C.~G\'eminard,
  {\em Phys. Rev. E} {\bf 84}, 061303 (2011).

\bibitem{Blanc2014} B.~Blanc, J.-C.~G{\'e}minard, and L.~A.~Pugnaloni,
  {\em The European Physical Journal E} {\bf 37}, 112 (2014).

\bibitem{Strogatz1993} S.~H.~Strogatz and I.~Stewart, {\em Scientific
  American} {\bf 269}, 68 (1993).

\bibitem{Jaeger1989} H.~M. Jaeger, C.~H.~Liu, and S.~R. Nagel, {\em
  Phys. Rev. Lett.} {\bf 62}, 40 (1989).

\bibitem{Xia2005} J. Xia, H. Gould, W. Klein, and J. B. Rundle,
  {\em Phys. Rev. Lett.} {\bf 95}, 248501 (2005).

\bibitem{Amundsen2012} D. S. Amundsen, J. Scheibert,
  K. Th{\o}gersen, J.~Tr\o mborg, and A. Malthe-S\o rensen, {\em
    Tribol. Lett.} {\bf 45}, 357 (2012).

\bibitem{LeBouil2014b} A.~Le~Bouil, A.~Amon, S.~McNamara, and
  J.~Crassous, {\em Phys. Rev. Lett.} {\bf 112}, 246001 (2014).

\bibitem{McNamara2016} S.~McNamara, J.~Crassous, and A.~Amon, {\em
  Phys. Rev. E} {\bf 94}, 022907 (2016).
\end{thebibliography}

\end{document}